\documentclass{article}
\usepackage[english]{babel}
\usepackage[cp1251]{inputenc}
\begin{document}
\begin{center}
\textbf{\Large{Applications of the operator $H\left( {\alpha
,\beta } \right)$ to the Humbert double hypergeometric functions}}\\
\medskip
\textbf{A. Hasanov}\\
\medskip
Institute of Mathematics, Uzbek Academy of Sciences,\\ 29, F.
Hodjaev street, Tashkent 100125, Uzbekistan\\
E-mail: anvarhasanov@yahoo.com
\end{center}
\medskip
\begin{abstract}

By making use of some techniques based upon certain inverse new
pairs of symbolic operators, the author investigate several
decomposition formulas associated with Humbert hypergeometric
functions $\Phi _1 $, $\Phi _2 $, $\Phi _3 $, $\Psi _1 $, $\Psi _2
$, $\Xi _1 $ and $\Xi _2 $. These operational representations are
constructed and applied in order to derive the corresponding
decomposition formulas. With the help of these inverse pairs of
symbolic operators, a total 34 decomposition formulas are found.
Euler type integrals, which are connected with Humbert's functions
are found.

\end{abstract}

\section{Introduction and definitions}

A great interest in the theory of hypergeometric functions (that
is, hypergeometric functions of several variables) is motivated
essentially by the fact that the solutions of many applied
problems involving (for example) partial differential equations
are obtainable with the help of such hypergeometric function (see,
for details, [21, p. 47]; see also the recent works [7-9, 16, 17]
and the references cited therein). For instance, the energy
absorbed by some nonferromagnetic conductor sphere included in an
internal magnetic field can be calculated with the help of such
functions [12, 15]. Hypergeometric functions of several variables
are used in physical and quantum chemical applications as well
[13, 19, 20]. Especially, many problems in gas dynamics lead to
solutions of degenerate second-order partial differential
equations, which are then solvable in terms of multiple
hypergeometric functions. Among examples, we can cite the problem
of adiabatic flat-parallel gas flow without whirlwind, the flow
problem of supersonic current from vessel with flat walls, and a
number of other problems connected with gas flow [2, 6]. We note
that Riemann's functions and fundamental solutions of the
degenerate second-order partial differential equations are
expressible by means of hypergeometric functions of several
variables [7-9]. In investigation of the boundary value problems
for these partial differential equations, we need decompositions
for hypergeometric functions of several variables in terms of
simpler hypergeometric functions of the Gauss and Humbert types.
    In particular, the Humbert functions $\Phi _1, \Phi _2, \Phi _3, \Psi _1, \Psi _2, \Xi _1 $
and $\Xi _2 $ in double variables, defined by ( [1, p. 126])
$$
\Phi _1 \left( {\alpha ,\beta ;\gamma ;x,y} \right) =
\displaystyle \sum\limits_{m,n = 0}^\infty  {} \frac{{\left(
\alpha \right)_{m + n} \left( \beta  \right)_m }}{{\left( \gamma
\right)_{m + n} m!n!}}x^m y^n ,\,\,\left| x \right| < 1,\,\,\left|
y \right| < \infty , \eqno (1.1)
$$
$$
\Phi _2 \left( {\beta _1 ,\beta _2 ;\gamma ;x,y} \right) =
\displaystyle \sum\limits_{m,n = 0}^\infty  {} \frac{{\left(
{\beta _1 } \right)_m \left( {\beta _2 } \right)_n }}{{\left(
\gamma \right)_{m + n} m!n!}}x^m y^n ,\,\,\left| x \right| <
\infty ,\,\,\left| y \right| < \infty ,\eqno (1.2)
$$
$$
\Phi _3 \left( {\beta ;\gamma ;x,y} \right) = \displaystyle
\sum\limits_{m,n = 0}^\infty  {} \frac{{\left( \beta \right)_m
}}{{\left( \gamma \right)_{m + n} m!n!}}x^m y^n ,\,\,\left| x
\right| < \infty ,\,\,\left| y \right| < \infty ,\eqno (1.3)
$$
$$
\Psi _1 \left( {\alpha ,\beta ;\gamma _1 ,\gamma _2 ;x,y} \right)
= \displaystyle \sum\limits_{m,n = 0}^\infty  {} \frac{{\left(
\alpha \right)_{m + n} \left( \beta  \right)_m }}{{\left( {\gamma
_1 } \right)_m \left( {\gamma _2 } \right)_n m!n!}}x^m y^n
,\,\,\left| x \right| < 1,\,\,\left| y \right| < \infty ,\eqno
(1.4)
$$
$$
\Psi _2 \left( {\alpha ;\gamma _1 ,\gamma _2 ;x,y} \right) =
\displaystyle \sum\limits_{m,n = 0}^\infty  {} \frac{{\left(
\alpha \right)_{m + n} }}{{\left( {\gamma _1 } \right)_m \left(
{\gamma _2 } \right)_n m!n!}}x^m y^n ,\,\,\left| x \right| <
\infty ,\,\,\left| y \right| < \infty ,\eqno (1.5)
$$
$$
\Xi _1 \left( {\alpha _1 ,\alpha _2 ,\beta ;\gamma ;x,y} \right) =
\displaystyle \sum\limits_{m,n = 0}^\infty  {} \frac{{\left(
{\alpha _1 } \right)_m \left( {\alpha _2 } \right)_n \left( \beta
\right)_m }}{{\left( \gamma  \right)_{m + n} m!n!}}x^m y^n
,\,\,\left| x \right| < 1,\,\,\left| y \right| < \infty ,\eqno
(1.6)
$$
$$
\Xi _2 \left( {\alpha ,\beta ;\gamma ;x,y} \right) = \displaystyle
\sum\limits_{m,n = 0}^\infty  {} \frac{{\left( \alpha \right)_m
\left( \beta  \right)_m }}{{\left( \gamma  \right)_{m + n}
m!n!}}x^m y^n ,\,\,\left| x \right| < 1,\,\,\left| y \right| <
\infty ,\eqno (1.7)
$$
and $\left( \alpha  \right)_m  = \Gamma \left( {\alpha  + m}
\right)/\Gamma \left( \alpha  \right)$ is the Pochhammer symbol.
For various multivariable hypergeometric functions including the
Lauricella multivariable functions $F_A^{\left( r \right)}$ ,
$F_B^{\left( r \right)} $, $F_C^{\left( r \right)}$ and
$F_D^{\left( r \right)}$, Hasanov and Srivastava [10, 11]
presented a number of decompositions formulas in terms of such
simpler hypergeometric functions as the Gauss and Appell
functions. The main object of this sequel to the works of Hasanov
and Srivastava [10, 11] is to show how some rather elementary
techniques based upon certain inverse pairs of symbolic operators
would lead us easily to several decomposition formulas associated
with Humbert's hypergeometric function $\Phi _1$, $\Phi _2$, $\Phi
_3$, $\Psi _1$, $\Psi _2$, $\Xi _1$ and $\Xi _2.$ Over six decades
ago, Burchnall and Chaundy [3, 4] and Chaundy [5] systematically
presented a number of expansion and decomposition formulas for
some double hypergeometric functions in series of simpler
hypergeometric functions. Their method is based upon the following
inverse pairs of symbolic operators:
$$
\nabla _{xy} \left( h \right): = \displaystyle \frac{{\Gamma
\left( h \right)\Gamma \left( {\delta _1 + \delta _2  + h}
\right)}}{{\Gamma \left( {\delta _1  + h} \right)\Gamma \left(
{\delta _2  + h} \right)}} = \sum\limits_{k = 0}^\infty  {}
\frac{{\left( { - \delta _1 } \right)_k \left( { - \delta _2 }
\right)_k }}{{\left( h \right)_k k!}}, \eqno (1.8)
$$
$$
\begin{array}{l}
\Delta _{xy} \left( h \right): = \displaystyle \frac{{\Gamma
\left( {\delta _1 + h} \right)\Gamma \left( {\delta _2  + h}
\right)}}{{\Gamma \left( h \right)\Gamma \left( {\delta _1  +
\delta _2  + h} \right)}} = \displaystyle \sum\limits_{k =
0}^\infty  {} \frac{{\left( { - \delta _1 } \right)_k \left( { -
\delta _2 }\right)_k }}{{\left( {1 - h - \delta _1  - \delta _2 } \right)_k k!}} \\
\,\,\,\,\,\,\,\,\,\,\,\,\,\,\,\,\,\,\,\,\,\,\, = \displaystyle
\sum\limits_{k = 0}^\infty  {} \frac{{\left( { - 1} \right)^k
\left( h \right)_{2k} \left( { - \delta _1 } \right)_k \left( { -
\delta _2 } \right)_k }}{{\left( {h + k - 1} \right)_k \left( {h +
\delta _1 }\right)_k \left( {h + \delta _2 } \right)_k k!}}, \\
\end{array} \eqno (1.9)
$$
and
$$
\begin{array}{l}
\nabla _{xy} \left( h \right)\Delta _{xy} \left( g \right):
=\displaystyle  \frac{{\Gamma \left( h \right) \Gamma \left(
{\delta _1  + \delta _2  + h} \right)}}{{\Gamma \left( {\delta _1
+ h} \right) \Gamma \left( {\delta _2  + h} \right)}}\frac{{\Gamma
\left( {\delta _1  + g} \right)\Gamma
\left( {\delta _2  + g} \right)}}{{\Gamma \left( g \right)\Gamma \left( {\delta _1  + \delta _2  + g} \right)}} \\
\,\,\,\,\,\,\,\,\,\,\,\,\,\,\,\,\,\,\,\,\,\,\,\,\,\,\,\,\,\,\,
=\displaystyle \sum\limits_{k = 0}^\infty  {} \displaystyle
\frac{{\left( {g - h} \right)_k \left( g \right)_{2k} \left( { -
\delta _1 } \right)_k \left( { - \delta _2 } \right)_k }}{{\left(
{g + k - 1} \right)_k \left( {g + \delta _1 }
\right)_k \left( {g + \delta _2 } \right)_k k!}} \\
\,\,\,\,\,\,\,\,\,\,\,\,\,\,\,\,\,\,\,\,\,\,\,\,\,\,\,\,\,\,\, =
\displaystyle \sum\limits_{k = 0}^\infty  {} \frac{{\left( {h - g}
\right)_k \left( { - \delta _1 } \right)_k \left( { - \delta _2 }
\right)_k }}{{\left( h \right)_k \left( {1 - g - \delta _1  -
\delta _2 } \right)_k k!}}, \,\,\,\,\,\,\left( {\delta _1 : =
\displaystyle  x\frac{\partial }{{\partial x}};\,\,\delta _2 : =
y\frac{\partial }{{\partial y}}} \right).
\end{array} \eqno (1.10)
$$
We introduce the following multivariable symbolic operators:
$$
H_{x_1 ,...,x_l } \left( {\alpha ,\beta } \right): = \displaystyle
\frac{{\Gamma \left( \beta  \right)\Gamma \left( {\alpha  + \delta
_1  +  \cdot \cdot  \cdot  + \delta _l } \right)}}{{\Gamma \left(
\alpha \right)\Gamma \left( {\beta  + \delta _1  +  \cdot  \cdot
\cdot  + \delta _l } \right)}} = \displaystyle \sum\limits_{k_1 ,
\cdot  \cdot \cdot ,k_l = 0}^\infty  {} \displaystyle
\frac{{\left( {\beta  - \alpha } \right)_{k_1  + \cdot  \cdot
\cdot  + k_l } \left( { - \delta _1 } \right)_{k_1 } \cdot  \cdot
\cdot \left( { - \delta _l } \right)_{k_l } }}{{\left( \beta
\right)_{k_1  + \cdot  \cdot \cdot  + k_l } k_1 ! \cdot  \cdot
\cdot k_l !}}\eqno (1.11)
$$
and
$$
\bar H_{x_1 ,...,x_l } \left( {\alpha ,\beta } \right): =
\displaystyle \frac{{\Gamma \left( \alpha  \right)\Gamma \left(
{\beta  + \delta _1  +  \cdot  \cdot  \cdot  + \delta _l }
\right)}}{{\Gamma \left( \beta  \right)\Gamma \left( {\alpha  +
\delta _1  +  \cdot  \cdot \cdot  + \delta _l } \right)}} =
\displaystyle \sum\limits_{k_1 ,...k_l  = 0}^\infty  {}
\displaystyle \frac{{\left( {\beta  - \alpha } \right)_{k_1  +
\cdot  \cdot \cdot  + k_l } \left( { - \delta _1 } \right)_{k_1 }
\cdot  \cdot \cdot \left( { - \delta _l } \right)_{k_l }
}}{{\left( {1 - \alpha - \delta _1  -  \cdot \cdot  \cdot  -
\delta _l } \right)_{k_1  + \cdot  \cdot  \cdot + k_l } k_1 !
\cdot  \cdot  \cdot k_l !}} \eqno (1.12)
$$
$$
\left( {\delta _j : = x_j \displaystyle \frac{\partial }{{\partial
x_j }},\,j = 1,...,l;\,\,l \in N: = \left\{ {1,2,3,...} \right\}}
\right).
$$

\section{Families of decompositions formulas for Humbert functions}

First of all, it is not difficult to derive the following
applications of the symbolic operators defined by (1.8) and (1.9):

$$
\Phi _1 \left( {\alpha ,\beta ;\gamma ;x,y} \right) =
\displaystyle H_{x,y} \left( {\alpha ,\varepsilon } \right)\Phi _1
\left( {\varepsilon ,\beta ;\gamma ;x,y} \right), \eqno (2.1)
$$
$$
\Phi _1 \left( {\alpha ,\beta ;\gamma ;x,y} \right) =
\displaystyle \bar H_{x,y} \left( {\varepsilon ,\alpha }
\right)\Phi _1 \left( {\varepsilon ,\beta ;\gamma ;x,y} \right),
\eqno (2.2)
$$
$$
\Phi _1 \left( {\alpha ,\beta ;\gamma ;x,y} \right) =
\displaystyle H_{x,y} \left( {\varepsilon ,\gamma } \right)\Phi _1
\left( {\alpha ,\beta ;\varepsilon ;x,y} \right), \eqno (2.3)
$$
$$
\Phi _1 \left( {\alpha ,\beta ;\gamma ;x,y} \right) =
\displaystyle H_{x,y} \left( {\alpha ,\gamma } \right)\left( {1 -
x} \right)^{ - \beta } e^y ,\eqno (2.4)
$$
$$
\left( {1 - x} \right)^{ - \beta } e^y  = \displaystyle \bar
H_{x,y} \left( {\alpha ,\gamma } \right)\Phi _1 \left( {\alpha
,\beta ;\gamma ;x,y} \right),\eqno (2.5)
$$
$$
\Phi _1 \left( {\alpha ,\beta ;\gamma ;x,y} \right) =
\displaystyle H_x \left( {\beta ,\varepsilon } \right)\Phi _1
\left( {\alpha ,\varepsilon ;\gamma ;x,y} \right),\eqno (2.6)
$$
$$
\Phi _1 \left( {\alpha ,\beta ;\gamma ;x,y} \right) =
\displaystyle \bar H_x \left( {\varepsilon ,\beta } \right)\Phi _1
\left( {\alpha ,\varepsilon ;\gamma ;x,y} \right),\eqno (2.7)
$$
$$
\Phi _2 \left( {\beta _1 ,\beta _2 ;\gamma ;x,y} \right)
=\displaystyle H_{x,y} \left( {\varepsilon ,\gamma } \right)\Phi
_2 \left( {\beta _1 ,\beta _2 ;\varepsilon ;x,y} \right),\eqno
(2.8)
$$
$$
\Phi _2 \left( {\beta _1 ,\beta _2 ;\gamma ;x,y} \right)
=\displaystyle H_x \left( {\beta _1 ,\varepsilon _1 } \right)\Phi
_2 \left( {\varepsilon _1 ,\beta _2 ;\gamma ;x,y} \right),\eqno
(2.9)
$$
$$
\Phi _2 \left( {\beta _1 ,\beta _2 ;\gamma ;x,y} \right) =
\displaystyle \bar H_x \left( {\varepsilon _1 ,\beta _1 }
\right)\Phi _2 \left( {\varepsilon _1 ,\beta _2 ;\gamma ;x,y}
\right),\eqno (2.10)
$$
$$
\Phi _2 \left( {\beta _1 ,\beta _2 ;\gamma ;x,y} \right) =
\displaystyle H_x \left( {\beta _1 ,\varepsilon _1 } \right)H_y
\left( {\beta _2 ,\varepsilon _2 } \right)\Phi _2 \left(
{\varepsilon _1 ,\varepsilon _2 ;\gamma ;x,y} \right),\eqno (2.11)
$$
$$
\Phi _2 \left( {\beta _1 ,\beta _2 ;\gamma ;x,y} \right) =
\displaystyle \bar H_x \left( {\varepsilon _1 ,\beta _1 }
\right)\bar H_y \left( {\varepsilon _2 ,\beta _2 } \right)\Phi _2
\left( {\varepsilon _1 ,\varepsilon _2 ;\gamma ;x,y} \right),\eqno
(2.12)
$$
$$
\Phi _3 \left( {\beta ;\gamma ;x,y} \right) = \displaystyle H_x
\left( {\beta ,\varepsilon } \right)\Phi _3 \left( {\varepsilon
;\gamma ;x,y} \right),\eqno (2.13)
$$
$$
\Phi _3 \left( {\beta ;\gamma ;x,y} \right) = \displaystyle \bar
H_x \left( {\varepsilon ,\beta } \right)\Phi _3 \left(
{\varepsilon ;\gamma ;x,y} \right),\eqno (2.14)
$$
$$
\Phi _3 \left( {\beta ;\gamma ;x,y} \right) =\displaystyle H_{x,y}
\left( {\varepsilon ,\gamma } \right)\Phi _3 \left( {\beta
;\varepsilon ;x,y} \right),\eqno (2.15)
$$
$$
\Psi _1 \left( {\alpha ,\beta ;\gamma _1 ,\gamma _2 ;x,y} \right)
= \displaystyle H_{x,y} \left( {\alpha ,\varepsilon } \right)\Psi
_1 \left( {\varepsilon ,\beta ;\gamma _1 ,\gamma _2 ;x,y}
\right),\eqno(2.16)
$$
$$
\Psi _1 \left( {\alpha ,\beta ;\gamma _1 ,\gamma _2 ;x,y} \right)
= \displaystyle H_x \left( {\beta ,\varepsilon } \right)\Psi _1
\left( {\alpha ,\varepsilon ;\gamma _1 ,\gamma _2 ;x,y}
\right),\eqno (2.17)
$$
$$
\Psi _1 \left( {\alpha ,\beta ;\gamma _1 ,\gamma _2 ;x,y} \right)
= \displaystyle \bar H_x \left( {\varepsilon ,\beta } \right)\Psi
_1 \left( {\alpha ,\varepsilon ;\gamma _1 ,\gamma _2 ;x,y}
\right),\eqno (2.18)
$$
$$
\Psi _1 \left( {\alpha ,\beta ;\gamma _1 ,\gamma _2 ;x,y} \right)
= \displaystyle H_x \left( {\beta ,\gamma _1 } \right)\left( {1 -
x} \right)^{ - \alpha } {}_1F_1 \left( {\alpha ,\gamma _2
;\frac{y}{{1 - x}}} \right),\eqno (2.19)
$$
$$
\left( {1 - x} \right)^{ - \alpha } {}_1F_1 \left( {\alpha ,\gamma
_2 ;\displaystyle \frac{y}{{1 - x}}} \right) = \bar H_x \left(
{\beta ,\gamma _1 } \right)\Psi _1 \left( {\alpha ,\beta ;\gamma
_1 ,\gamma _2 ;x,y} \right),\eqno (2.20)
$$
$$
\Psi _1 \left( {\alpha ,\beta ;\gamma _1 ,\gamma _2 ;x,y} \right)
= \displaystyle H_y \left( {\varepsilon ,\gamma _2 } \right)\Psi
_1 \left( {\alpha ,\beta ;\gamma _1 ,\varepsilon ;x,y}
\right),\eqno (2.21)
$$
$$
\Psi _2 \left( {\alpha ;\gamma _1 ,\gamma _2 ;x,y} \right)
=\displaystyle H_{x,y} \left( {\alpha ,\varepsilon } \right)\Psi
_2 \left( {\varepsilon ;\gamma _1 ,\gamma _2 ;x,y} \right),\eqno
(2.22)
$$
$$
\Psi _2 \left( {\alpha ;\gamma _1 ,\gamma _2 ;x,y} \right)
=\displaystyle \bar H_{x,y} \left( {\varepsilon ,\alpha }
\right)\Psi _2 \left( {\varepsilon ;\gamma _1 ,\gamma _2 ;x,y}
\right),\eqno (2.23)
$$
$$
\Psi _2 \left( {\alpha ;\gamma _1 ,\gamma _2 ;x,y} \right) =
\displaystyle H_x \left( {\varepsilon _1 ,\gamma _1 } \right)\Psi
_2 \left( {\alpha ;\varepsilon _1 ,\gamma _2 ;x,y} \right),\eqno
(2.24)
$$
$$
\Psi _2 \left( {\alpha ;\gamma _1 ,\gamma _2 ;x,y} \right) =
\displaystyle H_y \left( {\varepsilon _2 ,\gamma _2 } \right)\Psi
_2 \left( {\alpha ;\gamma _1 ,\varepsilon _2 ;x,y} \right),\eqno
(2.25)
$$
$$
\Psi _2 \left( {\alpha ;\gamma _1 ,\gamma _2 ;x,y} \right)
=\displaystyle H_x \left( {\varepsilon _1 ,\gamma _1 } \right)H_y
\left( {\varepsilon _2 ,\gamma _2 } \right)\Psi _2 \left( {\alpha
;\varepsilon _1 ,\varepsilon _2 ;x,y} \right),\eqno (2.26)
$$
$$
\Xi _1 \left( {\alpha _1 ,\alpha _2 ,\beta ;\gamma ;x,y} \right) =
\displaystyle H_x \left( {\alpha _1 ,\varepsilon _1 } \right)H_y
\left( {\alpha _2 ,\varepsilon _2 } \right)\Xi _1 \left(
{\varepsilon _1 ,\varepsilon _2 ,\beta ;\gamma ;x,y} \right),\eqno
(2.27)
$$
$$
\Xi _1 \left( {\alpha _1 ,\alpha _2 ,\beta ;\gamma ;x,y} \right)
=\displaystyle \bar H_x \left( {\varepsilon _1 ,\alpha _1 }
\right)\bar H_y \left( {\varepsilon _2 ,\alpha _2 } \right)\Xi _1
\left( {\varepsilon _1 ,\varepsilon _2 ,\beta ;\gamma ;x,y}
\right),\eqno (2.28)
$$
$$
\Xi _1 \left( {\alpha _1 ,\alpha _2 ,\beta ;\gamma ;x,y} \right)
=\displaystyle H_x \left( {\beta ,\varepsilon } \right)\Xi _1
\left( {\alpha _1 ,\alpha _2 ,\varepsilon ;\gamma ;x,y} \right),
\eqno (2.29)
$$
$$
\Xi _1 \left( {\alpha _1 ,\alpha _2 ,\beta ;\gamma ;x,y} \right)
=\displaystyle H_{x,y} \left( {\varepsilon ,\gamma } \right)\Xi _1
\left( {\alpha _1 ,\alpha _2 ,\beta ;\varepsilon ;x,y} \right),
\eqno (2.30)
$$
$$
\Xi _2 \left( {\alpha ,\beta ;\gamma ;x,y} \right) =\displaystyle
H_x \left( {\alpha ,\varepsilon _1 } \right)\Xi _2 \left(
{\varepsilon _1 ,\beta ;\gamma ;x,y} \right),\eqno (2.31)
$$
$$
\Xi _2 \left( {\alpha ,\beta ;\gamma ;x,y} \right) =\displaystyle
\bar H_x \left( {\varepsilon _1 ,\alpha } \right)\Xi _2 \left(
{\varepsilon _1 ,\beta ;\gamma ;x,y} \right),\eqno (2.32)
$$
$$
\Xi _2 \left( {\alpha ,\beta ;\gamma ;x,y} \right) =\displaystyle
H_x \left( {\beta ,\varepsilon _2 } \right)\Xi _2 \left( {\alpha
,\varepsilon _2 ;\gamma ;x,y} \right),\eqno (2.33)
$$
$$
\Xi _2 \left( {\alpha ,\beta ;\gamma ;x,y} \right) = \displaystyle
\bar H_x \left( {\varepsilon _2 ,\beta } \right)\Xi _2 \left(
{\alpha ,\varepsilon _2 ;\gamma ;x,y} \right),\eqno (2.34)
$$
$$
\Xi _2 \left( {\alpha ,\beta ;\gamma ;x,y} \right) =\displaystyle
H_{x,y} \left( {\varepsilon ,\gamma } \right)\Xi _2 \left( {\alpha
,\beta ;\varepsilon ;x,y} \right),\eqno (2.35)
$$
In view of the know Mellin-Barnes contour integral representations
for the Humbert functions $\Phi _1,$ $\Phi _2,$ $\Phi _3$, $\Psi
_1,$ $\Psi _2,$ $\Xi _1$ and $\Xi _2$, it is not difficult to give
alternative proofs of the operator identities (2.1) to (2.35)
above by using the Mellin and the inverse Mellin transformations
(see, for example, [14]). The details involved in these
alternative derivations of the operator identities (2.1) to (2.35)
are being omitted here. By virtue of the derivative formulas for
the Humbert functions, and also of some standard properties of
hypergeometric functions, we find each of the following
decomposition formulas for the Humbert $\Phi _1$, $\Phi _2$, $\Phi
_3$, $\Psi _1,$ $\Psi _2$, $\Xi _1$ and $\Xi _2$ in double
variables:
$$
\Phi _1 \left( {\alpha ,\beta ;\gamma ;x,y} \right) =
\displaystyle \sum\limits_{i,j = 0}^\infty  {} \frac{{\left( { -
1} \right)^{i + j} \left( {\varepsilon  - \alpha } \right)_{i + j}
\left( \beta \right)_i }}{{\left( \gamma  \right)_{i + j}
i!j!}}x^i y^j \Phi _1 \left( {\varepsilon  + i + j,\beta  +
i;\gamma  + i + j;x,y} \right),\eqno (2.36)
$$
$$
\Phi _1 \left( {\alpha ,\beta ;\gamma ;x,y} \right) =
\displaystyle \sum\limits_{i,j = 0}^\infty  {} \frac{{\left(
{\alpha  - \varepsilon } \right)_{i + j} \left( \beta  \right)_i
}}{{\left( \gamma  \right)_{i + j} i!j!}}x^i y^j \Phi _1 \left(
{\varepsilon ,\beta  + i;\gamma  + i + j;x,y} \right),\eqno (2.37)
$$
$$
\Phi _1 \left( {\alpha ,\beta ;\gamma ;x,y} \right) =
\displaystyle \sum\limits_{i,j = 0}^\infty  {} \frac{{\left( { -
1} \right)^{i + j} \left( \alpha  \right)_{i + j} \left( {\gamma
- \varepsilon } \right)_{i + j} \left( \beta  \right)_i }}{{\left(
\gamma \right)_{i + j} \left( \varepsilon  \right)_{i + j}
i!j!}}x^i y^j \Phi _1 \left( {\alpha  + i + j,\beta  +
i;\varepsilon  + i + j;x,y} \right),\eqno (2.38)
$$
$$
\Phi _1 \left( {\alpha ,\beta ;\gamma ;x,y} \right) =
\displaystyle e^y \left( {1 - x} \right)^{ - \beta } \Phi _1
\left( {\gamma  - \alpha ,\beta ;\gamma ;\frac{x}{{x - 1}}, - y}
\right),\eqno (2.39)
$$
$$
\left( {1 - x} \right)^{ - \beta } e^y  = \displaystyle
\sum\limits_{i,j = 0}^\infty  {} \frac{{\left( { - 1} \right)^{i +
j} \left( {\alpha - \gamma } \right)_{i + j} \left( \beta
\right)_i }}{{\left( \gamma  \right)_{i + j} i!j!}}x^i y^j \Phi _1
\left( {\alpha  + i + j,\beta  + i;\gamma  + i + j;x,y}
\right),\eqno (2.40)
$$
$$
\Phi _1 \left( {\alpha ,\beta ;\gamma ;x,y} \right) =
\displaystyle \sum\limits_{i = 0}^\infty  {} \frac{{\left( { - 1}
\right)^i \left( \alpha  \right)_i \left( {\varepsilon  - \beta }
\right)_i }}{{\left( \gamma  \right)_i i!}}x^i \Phi _1 \left(
{\alpha  + i,\varepsilon  + i;\gamma  + i;x,y} \right),\eqno
(2.41)
$$
$$
\Phi _1 \left( {\alpha ,\beta ;\gamma ;x,y} \right) =
\displaystyle \sum\limits_{i = 0}^\infty  {} \frac{{\left( \alpha
\right)_i \left( {\beta  - \varepsilon } \right)_i }}{{\left(
\gamma \right)_i i!}}x^i \Phi _1 \left( {\alpha  + i,\varepsilon
;\gamma + i;x,y} \right),\eqno (2.42)
$$
$$
\Phi _2 \left( {\beta _1 ,\beta _2 ;\gamma ;x,y} \right)
=\displaystyle \sum\limits_{i,j = 0}^\infty  {} \frac{{\left( { -
1} \right)^{i + j} \left( {\gamma  - \varepsilon } \right)_{i + j}
\left( {\beta _1 } \right)_i \left( {\beta _2 } \right)_j
}}{{\left( \gamma \right)_{i + j} \left( \varepsilon  \right)_{i +
j} i!j!}}x^i y^j \Phi _2 \left( {\beta _1  + i,\beta _2  +
j;\varepsilon  + i + j;x,y} \right), \eqno (2.43)
$$
$$
\Phi _2 \left( {\beta _1 ,\beta _2 ;\gamma ;x,y} \right)
=\displaystyle \sum\limits_{i = 0}^\infty  {} \frac{{\left( { - 1}
\right)^i \left( {\varepsilon _1  - \beta _1 } \right)_i
}}{{\left( \gamma \right)_i i!}}x^i \Phi _2 \left( {\varepsilon _1
+ i,\beta _2 ;\gamma  + i;x,y} \right), \eqno (2.44)
$$
$$
\Phi _2 \left( {\beta _1 ,\beta _2 ;\gamma ;x,y} \right)
=\displaystyle \sum\limits_{i = 0}^\infty  {} \frac{{\left( {\beta
_1  - \varepsilon _1 } \right)_i }}{{\left( \gamma  \right)_i
i!}}x^i \Phi _2 \left( {\varepsilon _1 ,\beta _2 ;\gamma  + i;x,y}
\right),\eqno (2.45)
$$
$$
\Phi _2 \left( {\beta _1 ,\beta _2 ;\gamma ;x,y} \right)
=\displaystyle \sum\limits_{i,j = 0}^\infty  {} \frac{{\left( { -
1} \right)^{i + j} \left( {\varepsilon _1  - \beta _1 } \right)_i
\left( {\varepsilon _2  - \beta _2 } \right)_j }}{{\left( \gamma
\right)_{i + j} i!j!}}x^i y^j \Phi _2 \left( {\varepsilon _1  +
i,\varepsilon _2  + j;\gamma  + i + j;x,y} \right),\eqno (2.46)
$$
$$
\Phi _2 \left( {\beta _1 ,\beta _2 ;\gamma ;x,y} \right)
=\displaystyle \sum\limits_{i,j = 0}^\infty  {} \frac{{\left(
{\beta _1  - \varepsilon _1 } \right)_i \left( {\beta _2  -
\varepsilon _2 } \right)_j }}{{\left( \gamma  \right)_{i + j}
i!j!}}x^i y^j \Phi _2 \left( {\varepsilon _1 ,\varepsilon _2
;\gamma  + i + j;x,y} \right), \eqno (2.4)
$$
$$
\Phi _3 \left( {\beta ;\gamma ;x,y} \right) =\displaystyle
\sum\limits_{i = 0}^\infty  {} \frac{{\left( { - 1} \right)^i
\left( {\varepsilon - \beta } \right)_i }}{{\left( \gamma
\right)_i i!}}x^i \Phi _3 \left( {\varepsilon  + i;\gamma  +
i;x,y} \right),\eqno (2.48)
$$
$$
\Phi _3 \left( {\beta ,\gamma ;x,y} \right) = \displaystyle
\sum\limits_{i = 0}^\infty  {} \frac{{\left( {\beta - \varepsilon
} \right)_i }}{{\left( \gamma  \right)_i i!}}x^i \Phi _3 \left(
{\varepsilon ,\gamma  + i;x,y} \right),\eqno (2.49)
$$
$$
\Phi _3 \left( {\beta ;\gamma ;x,y} \right) = \displaystyle
\sum\limits_{i,j = 0}^\infty  {} \frac{{\left( { - 1} \right)^{i +
j} \left( {\gamma - \varepsilon } \right)_{i + j} \left( \beta
\right)_i }}{{\left( \gamma  \right)_{i + j} \left( \varepsilon
\right)_{i + j} i!j!}}x^i y^j \Phi _3 \left( {\beta +
i;\varepsilon  + i + j;x,y} \right),\eqno (2.50)
$$
$$
\Psi _1 \left( {\alpha ,\beta ;\gamma _1 ,\gamma _2 ;x,y} \right)
= \displaystyle \sum\limits_{i,j = 0}^\infty  {} \frac{{\left( { -
1} \right)^{i + j} \left( {\varepsilon  - \alpha } \right)_{i + j}
\left( \beta \right)_i }}{{\left( {\gamma _1 } \right)_i \left(
{\gamma _2 } \right)_j i!j!}}x^i y^j \Psi _1 \left( {\varepsilon +
i + j,\beta + i;\gamma _1  + i,\gamma _2  + j;x,y} \right),\eqno
(2.51)
$$
$$
\Psi _1 \left( {\alpha ,\beta ;\gamma _1 ,\gamma _2 ;x,y} \right)
= \displaystyle \sum\limits_{i = 0}^\infty  {} \frac{{\left( { -
1} \right)^i \left( \alpha  \right)_i \left( {\varepsilon  - \beta
} \right)_i }}{{\left( {\gamma _1 } \right)_i i!}}x^i \Psi _1
\left( {\alpha + i,\varepsilon  + i;\gamma _1  + i,\gamma _2 ;x,y}
\right),\eqno (2.52)
$$
$$
\Psi _1 \left( {\alpha ,\beta ;\gamma _1 ,\gamma _2 ;x,y} \right)
= \displaystyle \sum\limits_{i = 0}^\infty  {} \frac{{\left(
\alpha \right)_i \left( {\beta  - \varepsilon } \right)_i
}}{{\left( {\gamma _1 } \right)_i i!}}x^i \Psi _1 \left( {\alpha
+ i,\varepsilon ;\gamma _1  + i,\gamma _2 ;x,y} \right),\eqno
(2.53)
$$
$$
\Psi _1 \left( {\alpha ,\beta ;\gamma _1 ,\gamma _2 ;x,y} \right)
=\displaystyle \left( {1 - x} \right)^{ - \alpha } \Psi _1 \left(
{\alpha ,\gamma _1  - \beta ;\gamma _1 ,\gamma _2 ;\frac{x}{{x -
1}},\frac{y}{{1 - x}}} \right),\eqno (2.54)
$$
$$
\left( {1 - x} \right)^{ - \alpha } {}_1F_1 \left( {\alpha ,\gamma
_2 ;\displaystyle \frac{y}{{1 - x}}} \right) = \displaystyle
\sum\limits_{i = 0}^\infty  {} \frac{{\left( \alpha \right)_i
\left( {\gamma _1  - \beta } \right)_i }}{{\left( {\gamma _1 }
\right)_i i!}}x^i \Psi _1 \left( {\alpha  + i,\beta ;\gamma _1  +
i,\gamma _2 ;x,y} \right),\eqno (2.55)
$$
$$
\Psi _1 \left( {\alpha ,\beta ;\gamma _1 ,\gamma _2 ;x,y} \right)
= \displaystyle \sum\limits_{i = 0}^\infty  {} \frac{{\left( { -
1} \right)^i \left( \alpha  \right)_i \left( {\gamma _2  -
\varepsilon } \right)_i }}{{\left( {\gamma _2 } \right)_i \left(
\varepsilon \right)_i i!}}y^i \Psi _1 \left( {\alpha  + i,\beta
;\gamma _1 ,\varepsilon  + i;x,y} \right),\eqno (2.56)
$$
$$
\Psi _2 \left( {\alpha ;\gamma _1 ,\gamma _2 ;x,y} \right)
=\displaystyle \sum\limits_{i,j = 0}^\infty  {} \frac{{\left( { -
1} \right)^{i + j} \left( {\varepsilon  - \alpha } \right)_{i + j}
}}{{\left( {\gamma _1 } \right)_i \left( {\gamma _2 } \right)_j
i!j!}}x^i y^j \Psi _2 \left( {\varepsilon  + i + j;\gamma _1  +
i,\gamma _2  + j;x,y} \right),\eqno (2.57)
$$
$$
\Psi _2 \left( {\alpha ;\gamma _1 ,\gamma _2 ;x,y} \right)
=\displaystyle \sum\limits_{i,j = 0}^\infty  {} \frac{{\left(
{\alpha  - \varepsilon } \right)_{i + j} }}{{\left( {\gamma _1 }
\right)_i \left( {\gamma _2 } \right)_j i!j!}}x^i y^j \Psi _2
\left( {\varepsilon ;\gamma _1  + i,\gamma _2  + j;x,y}
\right),\eqno (2.58)
$$
$$
\Psi _2 \left( {\alpha ;\gamma _1 ,\gamma _2 ;x,y} \right)
=\displaystyle \sum\limits_{i = 0}^\infty  {} \frac{{\left( { - 1}
\right)^i \left( \alpha  \right)_i \left( {\gamma _1  -
\varepsilon _1 } \right)_i }}{{\left( {\gamma _1 } \right)_i
\left( {\varepsilon _1 } \right)_i i!}}x^i \Psi _2 \left( {\alpha
+ i;\varepsilon _1  + i,\gamma _2 ;x,y} \right),\eqno (2.59)
$$
$$
\Psi _2 \left( {\alpha ;\gamma _1 ,\gamma _2 ;x,y} \right)
=\displaystyle \sum\limits_{i = 0}^\infty  {} \frac{{\left( { - 1}
\right)^i \left( \alpha  \right)_i \left( {\gamma _2  -
\varepsilon _2 } \right)_i }}{{\left( {\gamma _2 } \right)_i
\left( {\varepsilon _2 } \right)_i i!}}y^i \Psi _2 \left( {\alpha
+ i;\gamma _1 ,\varepsilon _2  + i;x,y} \right),\eqno (2.60)
$$
$$
\begin{array}{l}
\Psi _2 \left( {\alpha ;\gamma _1 ,\gamma _2 ;x,y} \right) \\
= \displaystyle \sum\limits_{i,j = 0}^\infty  {} \frac{{\left( { -
1} \right)^{i + j} \left( \alpha  \right)_{i + j} \left( {\gamma
_1  - \varepsilon _1 } \right)_i \left( {\gamma _2  - \varepsilon
_2 } \right)_j }}{{\left( {\gamma _1 } \right)_i \left( {\gamma _2
} \right)_j \left( {\varepsilon _1 } \right)_i \left( {\varepsilon
_2 } \right)_j i!j!}}x^i y^j \Psi _2 \left( {\alpha  + i + j;
\varepsilon _1  + i,\varepsilon _2  + j;x,y} \right), \\
\end{array}\eqno (2.61)
$$
$$
\begin{array}{l}
\Xi _1 \left( {\alpha _1 ,\alpha _2 ,\beta ;\gamma ;x,y} \right) =  \\
\displaystyle \sum\limits_{i,j = 0}^\infty  {} \frac{{\left( { - 1} \right)^{i + j}
\left( \beta  \right)_i \left( {\varepsilon _1  - \alpha _1 }
\right)_i \left( {\varepsilon _2  - \alpha _2 } \right)_j }}{{\left(
\gamma  \right)_{i + j} i!j!}}x^i y^j \Xi _1 \left( {\varepsilon _1  + i,\varepsilon _2  + j,\beta  + i;
\gamma  + i + j;x,y} \right), \\
\end{array}
\eqno (2.62)
$$
$$
\Xi _1 \left( {\alpha _1 ,\alpha _2 ,\beta ;\gamma ;x,y} \right) =
\displaystyle \sum\limits_{i,j = 0}^\infty  {} \frac{{\left( \beta
\right)_i \left( {\alpha _1  - \varepsilon _1 } \right)_i \left(
{\alpha _2 - \varepsilon _2 } \right)_j }}{{\left( \gamma
\right)_{i + j} i!j!}}x^i y^j \Xi _1 \left( {\varepsilon _1
,\varepsilon _2 ,\beta + i;\gamma  + i + j;x,y} \right), \eqno
(2.63)
$$
$$
\Xi _1 \left( {\alpha _1 ,\alpha _2 ,\beta ;\gamma ;x,y} \right) =
\displaystyle \sum\limits_{i = 0}^\infty  {} \frac{{\left( { - 1}
\right)^i \left( {\alpha _1 } \right)_i \left( {\varepsilon  -
\beta } \right)_i }}{{\left( \gamma  \right)_i i!}}x^i \Xi _1
\left( {\alpha _1  + i,\alpha _2 ,\varepsilon  + i;\gamma  +
i;x,y} \right), \eqno (2.64)
$$
$$
\begin{array}{l}
\Xi _1 \left( {\alpha _1 ,\alpha _2 ,\beta ;\gamma ;x,y} \right) \\
= \displaystyle \sum\limits_{i,j = 0}^\infty  {} \frac{{\left( { -
1} \right)^{i + j} \left( {\gamma  - \varepsilon } \right)_{i + j}
\left( {\alpha _1 } \right)_i \left( {\alpha _2 } \right)_j \left(
\beta  \right)_i }}{{\left( \gamma  \right)_{i + j} \left(
\varepsilon  \right)_{i + j} i!j!}}x^i y^j \Xi _1 \left( {\alpha
_1  + i,\alpha _2  + j,\beta  + i;
\varepsilon  + i + j;x,y} \right), \\
\end{array}\eqno (2.65)
$$
$$
\Xi _2 \left( {\alpha ,\beta ;\gamma ;x,y} \right) = \displaystyle
\sum\limits_{i = 0}^\infty  {} \frac{{\left( { - 1} \right)^i
\left( {\varepsilon _1  - \alpha } \right)_i \left( \beta
\right)_i }}{{\left( \gamma  \right)_i i!}}x^i \Xi _2 \left(
{\varepsilon _1  + i,\beta  + i;\gamma  + i;x,y} \right),\eqno
(2.66)
$$
$$
\Xi _2 \left( {\alpha ,\beta ;\gamma ;x,y} \right) = \displaystyle
\sum\limits_{i = 0}^\infty  {} \frac{{\left( {\alpha - \varepsilon
_1 } \right)_i \left( \beta  \right)_i }}{{\left( \gamma
\right)_i i!}}x^i \Xi _2 \left( {\varepsilon _1 ,\beta  + i;\gamma
+ i;x,y} \right),\eqno (2.67)
$$
$$
\Xi _2 \left( {\alpha ,\beta ;\gamma ;x,y} \right) = \displaystyle
\sum\limits_{i = 0}^\infty  {} \frac{{\left( { - 1} \right)^i
\left( \alpha  \right)_i \left( {\varepsilon _2  - \beta }
\right)_i }}{{\left( \gamma  \right)_i i!}}x^i \Xi _2 \left(
{\alpha  + i,\varepsilon _2  + i;\gamma  + i;x,y} \right),\eqno
(2.68)
$$
$$\Xi _2 \left( {\alpha ,\beta ;\gamma ;x,y} \right) =
\displaystyle \sum\limits_{i = 0}^\infty  {} \frac{{\left( \alpha
\right)_i \left( {\beta  - \varepsilon _2 } \right)_i }}{{\left(
\gamma \right)_i i!}}x^i \Xi _2 \left( {\alpha  + i,\varepsilon _2
;\gamma  + i;x,y} \right),\eqno (2.69)
$$
$$
\Xi _2 \left( {\alpha ,\beta ;\gamma ;x,y} \right) = \displaystyle
\sum\limits_{i,j = 0}^\infty  {} \frac{{\left( { - 1} \right)^{i +
j} \left( {\gamma  - \varepsilon } \right)_{i + j} \left( \alpha
\right)_i \left( \beta  \right)_i }}{{\left( \gamma \right)_{i +
j} \left( \varepsilon  \right)_{i + j} i!j!}}x^i y^j \Xi _2 \left(
{\alpha  + i,\beta  + i;\varepsilon  + i + j;x,y} \right).\eqno
(2.70)
$$
Our operational derivations of the decomposition formulas (2.15)
to (2.26) would indeed run parallel to those presented in the
earlier works, which we have already cited in the preceding
sections. In addition to the various operator expressions, we also
make use of the following operator identities [18, p. 93]:
$$
\left( {\delta  + \alpha } \right)_n \left\{ {f\left( \xi \right)}
\right\} =\displaystyle \xi ^{1 - \alpha } \frac{{d^n }}{{d\xi ^n
}}\left\{ {\xi ^{\alpha  + n - 1} f\left( \xi  \right)}
\right\}\eqno (2.71)
$$
$$
\left( {\delta : = \displaystyle \xi \frac{d}{{d\xi }};\,\,\alpha
\in C;\,\,n \in N_0 : =  N \cup \left\{ 0 \right\};\,\,N: =
\left\{ {1,2,3,...} \right\}} \right)
$$
and
$$
\left( { - \delta } \right)_n \left\{ {f\left( \xi  \right)}
\right\} = \left( { - 1} \right)^n \xi ^n \frac{{d^n }}{{d\xi ^n
}}\left\{ {f\left( \xi  \right)} \right\},\,\,\left( {\delta : =
\xi \frac{d}{{d\xi }};\,\,n \in N_0 } \right)\eqno (2.72)
$$
for every analytic function $f\left( \xi  \right)$. Many other
analogous decomposition formulas can similarly be derived for the
Humbert functions $\Phi _1 $, $\Phi _2$, $\Phi _3$, $\Psi _1$,
$\Psi _2$, $\Xi _1$ and $\Xi _2$, but with various different
parametric constraints.

\section{Demonstrations of some the decompositions formulas for Humbert functions}

The various decomposition formulas for the Humbert functions $\Phi
_1$, $\Phi _2$, $\Phi _3$, $\Psi _1$, $\Psi _2$, $\Xi _1$ and $\Xi
_2$ in double variables (which are presented here and in other
places in the previously cited literature) can be proven fairly
simply by suitably applying superposition of the inverse pairs of
symbolic operators introduced in Section 1. As an example, we
shall briefly indicate the proof of the decomposition formula
(2.36). For the double variable Humbert function $\Phi _1$, it is
not difficult to show from (2.1) that
$$
\Phi _1 \left( {\alpha ,\beta ;\gamma ;x,y} \right) =
\displaystyle \sum\limits_{i,j = 0}^\infty  {} \frac{{\left(
{\varepsilon  - \alpha } \right)_{i + j} \left( { - \delta _1 }
\right)_i \left( { - \delta _2 } \right)_j }}{{\left( \varepsilon
\right)_{i + j} i!j!}}\Phi _1 \left( {\varepsilon ,\beta ;\gamma
;x,y} \right),\eqno (3.1)
$$
$$
\left( {\delta _1 : = \displaystyle x\frac{\partial }{{\partial
x}};\,\,\delta _2 : = y\frac{\partial }{{\partial y}};} \right).
$$
Furthermore, by a straightforward computation, we have
$$
\left( { - \delta _1 } \right)_i \Phi _1 \left( {\varepsilon
,\beta ;\gamma ;x,y} \right) = \displaystyle \left( { - 1}
\right)^i x^i \frac{{\left( \varepsilon  \right)_i \left( \beta
\right)_i }}{{\left( \gamma \right)_i }}\Phi _1 \left(
{\varepsilon  + i,\beta  + i;\gamma  + i;x,y} \right),\eqno (3.2)
$$
and
$$
\left( { - \delta _2 } \right)_j \left( { - \delta _1 } \right)_i
\Phi _1 \left( {\varepsilon ,\beta ;\gamma ;x,y} \right) = \left(
{ - 1} \right)^{i + j} \frac{{\left( \varepsilon \right)_{i + j}
\left( \beta  \right)_i }}{{\left( \gamma \right)_{i + j} }}x^i
y^j \Phi _1 \left( {\varepsilon  + i + j,\beta  + i;\gamma  + i +
j;x,y} \right), \eqno (3.3)
$$
Upon substituting from (3.3) into (3.1), we finally arrive at the
decomposition formula (2.36).

\section{Integral representations via decomposition formulas}

Here in this section, we observe that several integral
representations of the Eulerian type can be deduced also from the
corresponding decomposition formulas of Section 2. For example,
using integral representations:
$$
\Phi _1 \left( {\alpha ,\beta ;\gamma ;x,y} \right) =
\displaystyle \frac{{\Gamma \left( \gamma \right)}}{{\Gamma \left(
\alpha \right)\Gamma \left( {\gamma  - \alpha }
\right)}}\int\limits_0^1 {} e^{y\xi } \xi ^{\alpha  - 1} \left( {1
- \xi } \right)^{\gamma - \alpha  - 1} \left( {1 - x\xi }
\right)^{ - \beta } d\xi ,\eqno (4.1)
$$

$$
\begin{array}{l}
\Phi _2 \left( {\beta _1 ,\beta _2 ;\gamma ;x,y} \right)
=\displaystyle \frac{{\Gamma \left( \gamma  \right)}}{{\Gamma
\left( {\beta _1 } \right)\Gamma \left( {\beta _2 } \right)\Gamma
\left( {\gamma  - \beta _1  - \beta _2 } \right)}} \\
\displaystyle \times \int\limits_0^1 {\int\limits_0^1 {} } e^{x\xi
+ y\left( {1 - \xi } \right)\eta } \xi ^{\beta _1  - 1} \eta
^{\beta _2  - 1} \left( {1 - \xi } \right)^{\gamma  - \beta _1  -
1} \left( {1 - \eta } \right)^{\gamma  -
\beta _1  - \beta _2  - 1} d\xi d\eta , \\
\end{array}\eqno
(4.2)
$$

$$
\begin{array}{l}
\Psi _1 \left( {\alpha ,\beta ;\gamma _1 ,\gamma _2 ;x,y} \right)
= \displaystyle \frac{{\Gamma \left( {\gamma _1 } \right)\Gamma
\left( {\gamma _2 } \right)}}{{\Gamma \left( \alpha \right)\Gamma
\left( \beta \right)\Gamma \left( {\gamma _1  - \beta }
\right)\Gamma \left( {\gamma _2  - \alpha } \right)}}\\
\displaystyle \times \int\limits_0^1 {\int\limits_0^1 {}
e^{\displaystyle \frac{{y\eta }}{{1 - x\xi }}} \xi ^{\beta  - 1}
\eta ^{\alpha - 1} \left( {1 - \xi } \right)^{\gamma _1  - \beta -
1} \left( {1 -
\eta }\right)^{\gamma _2  - \alpha  - 1} \left( {1 - x\xi } \right)^{ - \alpha } d\xi d\eta ,}  \\
\end{array}\eqno (4.3)
$$

$$
\begin{array}{l}
\Xi _1 \left( {\alpha _1 ,\alpha _2 ,\beta ;\gamma ;x,y} \right) =
\displaystyle \frac{{\Gamma \left( \gamma  \right)}}{{\Gamma
\left( {\alpha _1 } \right)\Gamma \left( {\alpha _2 }
\right)\Gamma \left( {\gamma  - \alpha _1  - \alpha _2 }
\right)}}\\
\displaystyle \times \int\limits_0^1 {\displaystyle
\int\limits_0^1 {}\displaystyle e^{y\left( {1 - \xi } \right)\eta
} \xi ^{\alpha _1  - 1} } \eta ^{\alpha _2 - 1} \left( {1 - \xi }
\right)^{\gamma  - \alpha _1  - 1} \left( {1 - \eta }
\right)^{\gamma  - \alpha _1  -
\alpha _2  - 1} \left( {1 - x\xi } \right)^{ - \beta } d\xi d\eta , \\
\end{array} \eqno (4.4)
$$

$$
\Xi _2 \left( {\alpha ,\beta ;\gamma ;x,y} \right) = \displaystyle
\frac{{\Gamma \left( \gamma  \right)}}{{\Gamma \left( \alpha
\right)\Gamma \left( {\gamma  - \alpha } \right)}}\int\limits_0^1
{} \xi ^{\alpha  - 1} \left( {1 - \xi } \right)^{\gamma  - \alpha
- 1} \left( {1 - x\xi } \right)^{ - \beta } {}_0F_1 \left[ {\gamma
- \alpha ;\left( {1 - \xi } \right)y} \right]d\xi \eqno (4.5)
$$
we have
$$
\begin{array}{l}
\Phi _1 \left( {\alpha ,\beta ;\gamma ;x,y} \right) \\
= \displaystyle \frac{{\Gamma \left( \gamma  \right)}}{{\Gamma
\left( \varepsilon \right)\Gamma \left( {\gamma  - \varepsilon }
\right)}}\int\limits_0^1 {} e^{y\xi } \xi ^{\varepsilon  - 1}
\left( {1 - \xi } \right)^{\gamma  - \varepsilon  - 1} \left( {1 -
x\xi } \right)^{ - \beta } \Phi _1 \left( {\varepsilon  - \alpha ,
\beta ;\varepsilon ;\frac{{x\xi }}{{x\xi  - 1}}, - y\xi } \right)d\xi , \\
{\mathop{\rm Re}\nolimits} \,\,\gamma  > {\mathop{\rm Re}\nolimits} \,\,\varepsilon  > 0, \\
\end{array}\eqno (4.6)
$$

$$
\begin{array}{l}
\Phi _1 \left( {\alpha ,\beta ;\gamma ;x,y} \right) =\displaystyle
\frac{{\Gamma \left( \gamma  \right)}}{{\Gamma \left( \varepsilon
\right)\Gamma \left( {\gamma  - \varepsilon }
\right)}}\\
\displaystyle \times \int\limits_0^1 {} e^{y\xi } \xi
^{\varepsilon - 1} \left( {1 - \xi } \right)^{\gamma  -
\varepsilon  - 1} \left( {1 - x\xi } \right)^{ - \beta } \Phi _1
\left( {\alpha  - \varepsilon , \beta ;\gamma  - \varepsilon
;\frac{{x\left( {1 - \xi } \right)}}{{1 - x\xi }},y\left( {1 -
\xi} \right)} \right)d\xi , \\
{\mathop{\rm Re}\nolimits} \,\,\gamma  > {\mathop{\rm Re}\nolimits} \,\,\varepsilon  > 0, \\
\end{array}\eqno (4.7)
$$

$$
\begin{array}{l}
\Phi _1 \left( {\alpha ,\beta ;\gamma ;x,y} \right) \\
=\displaystyle \frac{{\Gamma \left( \gamma  \right)}}{{\Gamma
\left( \alpha  \right)\Gamma \left( {\gamma  - \varepsilon }
\right)\Gamma \left( {\varepsilon  - \alpha }
\right)}}\int\limits_0^1 {\int\limits_0^1 {} } e^{y\xi \eta } \xi
^{\varepsilon  - 1} \eta ^{\alpha  - 1} \left( {1 - \xi }
\right)^{\gamma  - \varepsilon  - 1} \left( {1 -
\eta } \right)^{\varepsilon  - \alpha  - 1} \left( {1 - x\xi \eta } \right)^{ - \beta } d\xi d\eta , \\
{\mathop{\rm Re}\nolimits} \,\,\gamma  > {\mathop{\rm Re}\nolimits} \,\,\varepsilon  > 0,
\,{\mathop{\rm Re}\nolimits} \,\,\varepsilon  > {\mathop{\rm Re}\nolimits} \,\,\alpha  > 0, \\
\end{array}\eqno (4.8)
$$

$$
\begin{array}{l}
\Phi _1 \left( {\alpha ,\beta ;\gamma ;x,y} \right) =
\displaystyle \frac{{\Gamma \left( \gamma  \right)}}{{\Gamma
\left( \varepsilon  \right)\Gamma
\left( {\alpha  - \varepsilon } \right)\Gamma \left( {\gamma  - \alpha } \right)}} \\
\displaystyle \times \int\limits_0^1 {\int\limits_0^1 {} }
e^{y\left( {\xi  + \eta  - \xi \eta } \right)} \xi ^{\varepsilon -
1} \eta ^{\alpha  - \varepsilon  - 1} \left( {1 - \xi }
\right)^{\gamma  - \varepsilon  - 1} \left( {1 - \eta }
\right)^{\gamma  - \alpha  - 1} \left( {1 - x\xi  - x\eta  + x\xi
\eta }\right)^{ - \beta } d\xi d\eta , \\
{\mathop{\rm Re}\nolimits} \,\,\gamma  > {\mathop{\rm Re}\nolimits} \,\,\alpha  > 0,\,{\mathop{\rm Re}
\nolimits} \,\,\alpha  > {\mathop{\rm Re}\nolimits} \,\,\varepsilon  > 0, \\
\end{array} \eqno (4.9)
$$

$$
\begin{array}{l}
\Phi _2 \left( {\beta _1 ,\beta _2 ;\gamma ;x,y} \right)
=\displaystyle \frac{{\Gamma \left( \varepsilon  \right)}}{{\Gamma
\left( {\beta _1 } \right)\Gamma \left( {\beta _2 } \right)\Gamma
\left( {\varepsilon  - \beta _1  - \beta _2 }
\right)}} \\
\displaystyle \times \int\limits_0^1 {\int\limits_0^1 {} } e^{x\xi
+ y\left( {1 - \xi } \right)\eta } \xi ^{\beta _1  - 1} \eta
^{\beta _2  - 1} \left( {1 - \xi } \right)^{\varepsilon  - \beta
_1  - 1} \left( {1 - \eta } \right)^{\varepsilon  - \beta _1  -
\beta _2  - 1} {}_1F_1 \left( {\gamma  - \varepsilon ;
\gamma ; - x\xi  - y\left( {1 - \xi } \right)\eta } \right)d\xi d\eta , \\
{\mathop{\rm Re}\nolimits} \,\,\left( {\varepsilon  - \beta _1  - \beta _2 }
\right) > 0,\,\,{\mathop{\rm Re}\nolimits} \,\,\beta _1  > 0,\,\,\,{\mathop{\rm Re}\nolimits} \,\,\beta _2  > 0, \\
\end{array}\eqno (4.10)
$$

$$
\begin{array}{l}
\Phi _2 \left( {\beta _1 ,\beta _2 ;\gamma ;x,y} \right) =
\displaystyle \frac{{\Gamma \left( \gamma  \right)}}{{\Gamma
\left( {\varepsilon _1 } \right)\Gamma
\left( {\beta _2 } \right)\Gamma \left( {\gamma  - \varepsilon _1  - \beta _2 } \right)}} \\
\displaystyle \times \int\limits_0^1 {\int\limits_0^1 {} } e^{x\xi
+ y\left( {1 - \xi } \right)\eta } \xi ^{\varepsilon _1  - 1} \eta
^{\beta _2  - 1} \left( {1 - \xi } \right)^{\gamma  - \varepsilon
_1  - 1} \left( {1 - \eta } \right)^{\gamma  - \varepsilon _1  -
\beta _2  - 1} {}_1F_1
\left( {\varepsilon _1  - \beta _1 ;\varepsilon _1 ; - x\xi } \right)d\xi d\eta , \\
{\mathop{\rm Re}\nolimits} \,\,\left( {\gamma  - \varepsilon _1  - \beta _2 }
\right) > 0,\,\,{\mathop{\rm Re}\nolimits} \,\,\varepsilon _1  > 0,\,\,\,{\mathop{\rm Re}
\nolimits} \,\,\beta _2  > 0, \\
\end{array}\eqno (4.11)
$$

$$
\begin{array}{l}
\Phi _2 \left( {\beta _1 ,\beta _2 ;\gamma ;x,y} \right) \\ =
\displaystyle \frac{{\Gamma \left( \gamma  \right)}}{{\Gamma
\left( {\varepsilon _1 } \right)\Gamma \left( {\beta _2 }
\right)\Gamma \left( {\gamma  - \varepsilon _1  - \beta _2 }
\right)}} \displaystyle \int\limits_0^1 {\int\limits_0^1 {} }
e^{x\xi + y\left( {1 - \xi } \right)\eta } \xi ^{\varepsilon _1 -
1} \eta ^{\beta _2  - 1} \left( {1 - \xi } \right)^{\gamma -
\varepsilon _1  - 1} \left( {1 - \eta } \right)^{\gamma  -
\varepsilon _1  - \beta _2  - 1}\\
\times {}_1F_1 \left( {\beta _1  - \varepsilon _1 ; \gamma  -
\varepsilon _1  - \beta _2 ;x\left( {1 - \xi }
\right)\left( {1 - \eta } \right)} \right)d\xi d\eta , \\
{\mathop{\rm Re}\nolimits} \,\,\left( {\gamma  - \varepsilon _1  - \beta _2 }
\right) > 0,\,\,{\mathop{\rm Re}\nolimits} \,\,\varepsilon _1  > 0,\,\,\,{\mathop{\rm Re}
\nolimits} \,\,\beta _2  > 0, \\
\end{array}\eqno (4.12)
$$

$$
\begin{array}{l}
\Phi _2 \left( {\beta _1 ,\beta _2 ;\gamma ;x,y} \right) \\
= \displaystyle \frac{{\Gamma \left( \gamma  \right)}}{{\Gamma
\left( {\varepsilon _1 } \right)\Gamma \left( {\varepsilon _2 }
\right)\Gamma \left( {\gamma  - \varepsilon _1  - \varepsilon _2 }
\right)}}\displaystyle \int\limits_0^1 {\int\limits_0^1 {} }
e^{x\xi  + y\left( {1 - \xi } \right)\eta } \xi ^{\varepsilon _1 -
1} \eta ^{\varepsilon _2  - 1} \left( {1 - \xi } \right)^{\gamma -
\varepsilon _1  - 1} \left( {1 - \eta } \right)^{\gamma  - \varepsilon _1  - \varepsilon _2  - 1}  \\
\times {}_1F_1 \left( {\varepsilon _1  - \beta _1 ;\varepsilon _1 ; - x\xi } \right){}_1F_1
\left( {\varepsilon _2  - \beta _2 ;\varepsilon _2 ; - y\left( {1 - \xi } \right)\eta } \right)d\xi d\eta , \\
{\mathop{\rm Re}\nolimits} \,\,\left( {\gamma  - \varepsilon _1  - \varepsilon _2 }
\right) > 0,\,\,{\mathop{\rm Re}\nolimits} \,\,\varepsilon _1  > 0,\,\,\,{\mathop{\rm Re}\nolimits}
\,\,\varepsilon _2  > 0, \\
\end{array}\eqno (4.13)
$$

$$
\begin{array}{l}
\Phi _2 \left( {\beta _1 ,\beta _2 ;\gamma ;x,y} \right) =
\displaystyle \frac{{\Gamma \left( \gamma \right)}}{{\Gamma \left(
{\varepsilon _1 } \right)\Gamma \left( {\varepsilon _2 }
\right)\Gamma \left( {\gamma  - \varepsilon _1  - \varepsilon _2 }
\right)}} \displaystyle \int\limits_0^1 {\int\limits_0^1 {} }
e^{x\xi  + y\left( {1 - \xi } \right)\eta }
\xi ^{\varepsilon _1  - 1} \eta ^{\varepsilon _2  - 1}  \\
\times \left( {1 - \xi } \right)^{\gamma  - \varepsilon _1  - 1} \left( {1 - \eta } \right)^{\gamma  -
\varepsilon _1  - \varepsilon _2  - 1} \Phi _2 \left( {\beta _1  - \varepsilon _1 ,\beta _2  -
\varepsilon _2 ;\gamma ;x\left( {1 - \xi } \right)\left( {1 - \eta } \right),y\left( {1 - \xi }
\right)\left( {1 - \eta } \right)} \right)d\xi d\eta , \\
{\mathop{\rm Re}\nolimits} \,\,\left( {\gamma  - \varepsilon _1  - \varepsilon _2 }
\right) > 0,\,\,{\mathop{\rm Re}\nolimits} \,\,\varepsilon _1  > 0,\,\,\,{\mathop{\rm Re}
\nolimits} \,\,\varepsilon _2  > 0, \\
\end{array} \eqno(4.14)
$$

$$
\begin{array}{l}
\Psi _1 \left( {\alpha ,\beta ;\gamma _1 ,\gamma _2 ;x,y} \right) = \displaystyle \frac{{\Gamma \left( {\gamma _1 }
\right)\Gamma \left( \varepsilon  \right)}}{{\Gamma \left( \alpha  \right)\Gamma
\left( \beta  \right)\Gamma \left( {\gamma _1  - \beta } \right)\Gamma \left( {\varepsilon  - \alpha } \right)}} \\
\displaystyle \times \int\limits_0^1 {\int\limits_0^1 {}
e^{\displaystyle \frac{{y\eta }}{{1 - x\xi }}} \xi ^{\beta  - 1}
\eta ^{\alpha - 1} \left( {1 - \xi } \right)^{\gamma _1  - \beta -
1} \left( {1 - \eta } \right)^{\varepsilon  - \alpha  - 1} \left(
{1 - x\xi } \right)^{ - \alpha } {}_1F_1 \left[ {\gamma _2  - \varepsilon ;
\gamma _2 ;\frac{y}{{x\xi  - 1}}} \right]d\xi d\eta } , \\
{\mathop{\rm Re}\nolimits} \,\,\gamma _1  > {\mathop{\rm Re}\nolimits}
\,\,\beta  > 0,\,\,{\mathop{\rm Re}\nolimits} \,\,\varepsilon  > \,\,{\mathop{\rm Re}\nolimits} \,\,\alpha  > 0, \\
\end{array}\eqno
(4.15)
$$

$$
\begin{array}{l}
\Xi _1 \left( {\alpha _1 ,\alpha _2 ,\beta ;\gamma ;x,y} \right) =
\displaystyle \frac{{\Gamma \left( \gamma  \right)}}{{\Gamma
\left( {\varepsilon _1 } \right)\Gamma \left( {\varepsilon _2 }
\right)\Gamma \left( {\gamma  - \varepsilon _1  - \varepsilon _2 }\right)}} \\
\times \displaystyle \int\limits_0^1 {\int\limits_0^1 {}
e^{y\left( {1 - \xi } \right)\eta } \xi ^{\varepsilon _1  - 1} }
\eta ^{\varepsilon _2 - 1} \left( {1 - \xi } \right)^{\gamma  -
\varepsilon _1  - 1} \left( {1 - \eta } \right)^{\gamma  -
\varepsilon _1  - \varepsilon _2  - 1}
\left( {1 - x\xi } \right)^{ - \beta }  \\
\displaystyle \times \Xi _1 \left( {\alpha _1  - \varepsilon _1
,\alpha _2  - \varepsilon _2 ,\beta ;\gamma  - \varepsilon _1  -
\varepsilon _2 ;\frac{{x\left( {1 - \xi } \right)\left( {1 - \eta
}\right)}}{{1 - x\xi }},y\left( {1 - \xi } \right)\left( {1 - \eta } \right)} \right)d\xi d\eta , \\
{\mathop{\rm Re}\nolimits} \,\,\left( {\gamma  - \varepsilon _1  - \varepsilon _2 }
\right) > 0,\,\,{\mathop{\rm Re}\nolimits} \,\,\varepsilon _1  > 0,\,\,\,{\mathop{\rm Re}
\nolimits} \,\,\varepsilon _2  > 0, \\
\end{array}\eqno(4.16)
$$

$$
\begin{array}{l}
\Xi _1 \left( {\alpha _1 ,\alpha _2 ,\beta ;\gamma ;x,y} \right) =
\displaystyle \frac{{\Gamma \left( \gamma  \right)}}{{\Gamma
\left( {\varepsilon _1 } \right)\Gamma \left( {\varepsilon _2 }
\right)\Gamma \left( {\gamma  - \varepsilon _1  - \varepsilon _2 }
\right)}}\displaystyle \int\limits_0^1 {\int\limits_0^1 {}
e^{y\left( {1 - \xi } \right)\eta } } \xi ^{\varepsilon _1  - 1} \eta ^{\varepsilon _2  - 1}  \\
\displaystyle \times \left( {1 - \xi } \right)^{\gamma  -
\varepsilon _1  - 1} \left( {1 - \eta } \right)^{\gamma  -
\varepsilon _1  - \varepsilon _2  - 1} F\left( {\alpha _1 ,\beta ;
\varepsilon _1 ;x\xi } \right){}_1F_1 \left( {\varepsilon _2  -
\alpha _2 , \varepsilon _2 ; - y\left( {1 - \xi } \right)\eta } \right)d\xi d\eta , \\
{\mathop{\rm Re}\nolimits} \,\,\left( {\gamma  - \varepsilon _1  -
\varepsilon _2 } \right) > 0,\,\,{\mathop{\rm Re}\nolimits}
\,\,\varepsilon _1  > 0,\,\,\,{\mathop{\rm Re}\nolimits}
\,\,\varepsilon _2  > 0, \\
\end{array}\eqno (4.17)
$$

$$
\begin{array}{l}
\Xi _2 \left( {\alpha ,\beta ;\gamma ;x,y} \right) \\
= \displaystyle \frac{{\Gamma \left( \gamma  \right)}}{{\Gamma
\left( {\varepsilon _1 } \right)\Gamma \left( {\gamma  -
\varepsilon _1 } \right)}}\displaystyle \int\limits_0^1 {} \xi
^{\varepsilon _1  - 1} \left( {1 - \xi } \right)^{\gamma  -
\varepsilon _1  - 1} F\left( {\alpha , \beta ;\varepsilon _1 ;x\xi
} \right){}_0F_1 \left( {\gamma  -
\varepsilon _1 ;y\left( {1 - \xi } \right)} \right)d\xi , \\
{\mathop{\rm Re}\nolimits} \,\,\gamma  > \,\,{\mathop{\rm Re}\nolimits} \,\,\varepsilon _1  > 0. \\
\end{array}\eqno
(4.18)
$$

$$
\begin{array}{l}
\Xi _2 \left( {\alpha ,\beta ;\gamma ;x,y} \right) = \displaystyle
\frac{{\Gamma \left( \gamma \right)}}{{\Gamma \left( \alpha
\right)\Gamma \left( {\varepsilon _1  - \alpha } \right)
\Gamma \left( {\gamma  - \varepsilon _1 } \right)}} \\
\displaystyle \int\limits_0^1 {\int\limits_0^1 {} } \xi
^{\varepsilon _1  - 1} \eta ^{\alpha  - 1} \left( {1 - \xi }
\right)^{\gamma  - \varepsilon _1  - 1} \left( {1 - \eta }
\right)^{\varepsilon _1  - \alpha  - 1} \left( {1 - x\xi \eta }
\right)^{ - \beta } {}_0F_1 \left( {\gamma  - \varepsilon _1
;y\left( {1 - \xi } \right)} \right)d\xi d\eta , \\
{\mathop{\rm Re}\nolimits} \,\,\gamma  > \,\,{\mathop{\rm Re}\nolimits} \,\,\varepsilon _1  >
\,\,{\mathop{\rm Re}\nolimits} \,\,\alpha  > 0, \\
\end{array}\eqno
(4.19)
$$

$$
\begin{array}{l}
\Xi _2 \left( {\alpha ,\beta ;\gamma ;x,y} \right) = \displaystyle
\frac{{\Gamma \left( \gamma  \right)}}{{\Gamma \left( \beta
\right)\Gamma \left( {\gamma  - \varepsilon _1 } \right)\Gamma
\left( {\varepsilon _1  - \beta } \right)}} \\
\displaystyle \times \int\limits_0^1 {\int\limits_0^1 {} } \xi
^{\varepsilon _1  - 1} \eta ^{\beta  - 1} \left( {1 - \xi }
\right)^{\gamma  - \varepsilon _1  - 1} \left( {1 - \eta }
\right)^{\varepsilon _1  - \beta  - 1} \left( {1 - x\xi \eta } \right)^{ - \alpha } {}_0F_1
\left( {\gamma  - \varepsilon _1 ;y\left( {1 - \xi } \right)} \right)d\xi d\eta , \\
{\mathop{\rm Re}\nolimits} \,\,\gamma  > \,\,{\mathop{\rm Re}\nolimits} \,\,\varepsilon _1  >
\,\,{\mathop{\rm Re}\nolimits} \,\,\beta  > 0. \\
\end{array}\eqno (4.20)
$$
\textbf{Remark.} Note, that mutually inverse operators (1.11) and
(1.12) can be applied to other multivariate hypergeometric
functions.

\end{document}